\documentclass[10pt, a4paper, twocolumn]{article}
\usepackage[backend=biber, style=numeric-comp, natbib,sorting=none,maxbibnames = 5,giveninits = true]{biblatex}
\DeclareNameAlias{author}{family-given}
\renewbibmacro{in:}{
  \ifentrytype{article}
    {}
    {\bibstring{in}%
     \printunit{\intitlepunct}}} 

\DeclareFieldFormat{pages}{#1} 
\usepackage{abstract}
\usepackage{authblk}
\usepackage{graphicx}
\usepackage{hyperref}
\usepackage{subcaption}
\usepackage{booktabs}
\usepackage{amsmath, amssymb}
\usepackage{caption} 

\usepackage{lipsum}
\usepackage{geometry}
\usepackage{enumitem}
 \usepackage{xurl}

\providecommand{\keywords}[1]
{
  \small	
  \textbf{\textit{Keywords---}} #1
}

\title{\textbf{\Large{COVID-19 and social media: Beyond polarization}}}

\author[*,a]{Giacomo De Nicola}
\author[b,a]{Victor H. Tuekam Mambou}
\author[a]{G{\"o}ran Kauermann}

\affil[a]{Department of Statistics, Ludwig Maximilian University of
  Munich, Germany}
\affil[b]{ifo Institute -- Leibniz Institute for Economic Research at the University of Munich, Germany}
\affil[*]{{corresponding author - giacomo.denicola@stat.uni-muenchen.de}}
\date{}                     
\newcommand{\abstractText}{\noindent
The COVID-19 pandemic brought upon a massive wave of disinformation, exacerbating polarization in the increasingly divided landscape of online discourse. In this context, popular social media users play a major role, as they have the ability to broadcast messages to large audiences and influence public opinion. In this paper, we make use of openly available data to study the behavior of popular users discussing the pandemic on Twitter. We tackle the issue from a network perspective, considering users as nodes and following relationships as directed edges. The resulting network structure is modeled by embedding the actors in a latent social space, where users closer to one another have a higher probability of following each other. The results suggest the existence of two distinct communities, which can be interpreted as ``generally pro'' and ``generally against'' vaccine mandates, corroborating existing evidence on the pervasiveness of echo chambers on the platform. By focusing on a number of notable users, such as politicians, activists, and news outlets, we further show that the two groups are not entirely homogeneous, and that not just the two poles are represented. To the contrary, the latent space captures an entire spectrum of beliefs between the two extremes, demonstrating that polarization, while present, is not the only driver of the network, and that more moderate, ``central'' users are key players in the discussion.

\vspace{7pt}
\keywords{Polarization, COVID-19, Network analysis, Twitter, Latent space models}
}

\usepackage{hyperref}
\hypersetup{colorlinks=true, urlcolor=blue, linkcolor=blue, citecolor=blue}

\begin{document}
\begin{refsection}

\twocolumn[
  \begin{@twocolumnfalse}
    \maketitle
    \begin{abstract}
      \abstractText
      \newline
      \newline
    \end{abstract}
    \paragraph{Significance Statement:}{Popular social media users play a major role in the COVID-19 infodemic, as they can influence public opinion through their massive reach. Using state-of-the-art statistical network modeling techniques, we embed popular Twitter users discussing the pandemic in a latent social space, producing a map of the COVID-19 social media universe. The results suggest the existence of two distinct communities, which respectively favor and oppose vaccine mandates, thus corroborating the presence of echo chamber effects on the platform. We further show that the two groups are not entirely homogeneous: instead, the social map describes an entire spectrum of beliefs between the two extremes, demonstrating that polarization is not the only relevant factor, and that moderate users are central to the discussion.}
\vspace{1cm}
  \end{@twocolumnfalse}
]


\section*{Introduction}

COVID-19 dramatically affected the lives of billions of people around the globe. Given its massive impact, the pandemic naturally assumed a central role in both private and public discourse, dominating the discussion on- and offline. Social media, in particular, has been extensively used to exchange pandemic-related information as well as disinformation, leading to what has been defined as an ``infodemic'' alongside the pandemic \cite{cinelli2020,zarocostas2020,gollust2020}. 
 This context saw the emergence of pandemic-related social media \textsl{elites}, accounts with a large number of followers that regularly discuss the pandemic and the issues surrounding it \cite{gallagher2021,molyneux2021}. These actors play a central role in public communication, as they can shape popular sentiment and public discourse and thus potentially influence political decision-making \cite{leader2021}. This is especially true in a setting characterized by increasing polarization and historically low trust in mainstream news, which allows politically and financially motivated actors to emerge \cite{finkel2020,fink2019,bridgman2020,donovan2020}. Because of this, understanding the role that popular social media users play and the ways in which they operate is crucial for tackling arising challenges in public communication \cite{velasquez2020}. 
 In this paper, we tackle this issue with the aim of drawing an explanatory map of the network of COVID-19 Twitter elites. We first identify users that are popular in the discussion related to the pandemic on Twitter, and go on to study their (directed) network, where an edge between two actors is present if one follows the other. To analyze the resulting network structure we make use of latent space models, which postulate that nodes in the network are embedded in a latent social space, where the probability for two actors to connect is inversely related to their distance within the space \cite{hoff2002}. We, in particular, make use of the latent cluster random effects model, which incorporates model-based clustering, allowing it to identify cohesive communities in the network, as well as additional nodal parameters to account for actor-specific heterogeneity in the propensity to form edges \cite{krivitsky2009}. The results suggest that the network can be partitioned into two macro-communities. By focusing on a number of notable users, such as politicians, activists, and news outlets, we show how the two communities can be interpreted as ``generally pro'' and ``generally against'' pandemic containment measures and vaccine mandates. This finding supports the extensive body of literature that demonstrates the existence of significant polarization on social media
 \cite{caldarelli2020role, conover2011political, garimella2017long, del2016echo, del2017mapping}. The central role of polarization has also been demonstrated for the specific case of pandemic-related online conversations, especially with respect to opinions on vaccination \cite{jiang2021,reiter2022polarization, steelfisher2021,cowan2021,monsted2022}. 
  However, our results also demonstrate how polarization, while prevalent, is far from being the only driver in the network. The continuous latent space enables us to see that substantial within-cluster heterogeneity is present: not all users in the two communities have the same opinions, and not just the two polar opposites are represented. On the contrary, a full spectrum of beliefs between the two poles is found. In particular, more radical users are found to be positioned towards the extremes of the latent space, while more moderate and neutral actors, such as health ministers and news outlets, are closer to the center. These central users thus occupy a uniquely powerful position, as they can act as a bridge between the two communities and thereby mitigate polarization. 
  In addition to these results, our analysis demonstrates how, by making use of latent space models, it is possible to accurately map the COVID-19 Twitter landscape by only modeling information on who follows whom within the elite network. This finding highlights the strength and the pervasiveness of echo chamber effects on the platform, and showcases the power of latent space network models for studying communication on social media.

\section*{Data and Methods}

\subsubsection*{Identifying the network of COVID-19 Twitter elites}

  Social media elites can be broadly understood as users with the ability to influence \cite{dubois2014}. The term typically refers to a group of highly influential and popular users with considerable reach who significantly impact conversations, trends, and narratives circulating on social media. These users often include celebrities, politicians, journalists, thought leaders and influencers, who have a large and engaged audience and are frequently retweeted, quoted and mentioned by others. While informative, this characterization is quite broad and does not indicate a unique way of identifying elites in practice. Operational definitions for empirical applications are often based on engagement metrics, such as the number of followers of each user, and engagement metrics, such as likes, shares, quotes, and replies. As our focus lies on analyzing the behavior of actors who actively engage in the discussion of the pandemic and that exert significant influence on the conversation, we here choose to identify elites as those who authored the most popular tweets, where the popularity of a tweet is given by the sum of its likes, replies and retweets (including quotes). Based on this characterization, we will therefore first need to identify popular tweets discussing COVID-19, and then relate those tweets to their authors. More motivation and details on this choice, as well as robustness checks, are included in the supplementary material.
For our study, we make use of the COVID-19 Twitter dataset published by Banda et al. \cite{banda2021}, which comprises IDs of tweets containing pandemic-related keywords from January 1, 2020, onward. These keywords were handpicked and continuously tracked to provide a global and real-time overview of the chatter related to the COVID-19 pandemic. The dataset was collected using Twitter's streaming API, which allows free access to a random 1\% sample of publicly available tweets in real-time \cite{twitter}.
At the time of the analysis, the entire dataset contains about 1.32 billion tweet IDs, representing both tweets and retweets in all languages, 340 million of which are unique (without retweets). 
Each tweet's creation time and language are also provided.
Using the tweet IDs, we are then able to recover additional information on the tweets, such the text, the author, and metrics such as likes and retweets counts. 

As a global platform, Twitter is host to speakers of many different languages, which induce the formation of largely separate communities. Since our goal is to map the latent space of COVID-19 elites, we choose to limit our analysis to a single language, as doing otherwise would return a fragmented map shaped mainly by language. In principle, it is possible to work with any single language, and we here opt for using tweets in German. The choice is motivated by the combination of two facts: Firstly, German is predominantly spoken by people from Germany, and to a smaller extent from Austria and parts of Switzerland, thereby guaranteeing a reasonable degree of geographical homogeneity. This prevents the estimated latent positions of the actors (and the resulting clusters) from being predominantly driven by their geographical locations. Secondly, German is used by a relevant proportion of the Twitter user base, allowing for a more than sufficient sample size. 
As the first COVID-19 vaccines started to be available to the public towards the very end of 2020, and given that one of the points we are most interested in investigating is attitude towards vaccination, we limit our sample to 2021 only, spanning from January 1 to December 31. Considering all tweets in German from 2021 results in a total of 1.51 million unique tweets from 184,406 accounts. The data, sketched in Table \ref{tab:data}, allows us to pinpoint popular users by looking at the authors of tweets with the highest interaction metrics. More specifically, we classify a user as elite if they authored a tweet that achieved a  popularity score of at least 2000, where we define popularity as the sum of likes, replies, and retweets (including quotes) gathered. This threshold results in 1024 popular tweets spanning all months of 2021, with each month represented by 53--156 tweets. 
Those 1024 tweets were produced by 372 users,  31.7\% of which were granted verified status by Twitter, meaning that the platform deemed them both authentic and of public interest \cite{edgerly2019}. In contrast, only 2.4\% of the user base in the initial sample was verified. This confirms that more notable accounts and public figures are, on average, more central to the discussion, as we would expect.
\begin{table}[t]
\small\sf\centering
\caption{Structure of the analyzed dataset. Only columns relevant to our study are displayed.\label{tab:data}}
\begin{tabular}{lllll}
\toprule
\textbf{{Tweet ID}} &\textbf{{Author}} & \textbf{{Likes}} & \textbf{Replies} &\textbf{{Retweets}}\\
\midrule
138712... &AnikaBlub &1,162&61&53\\
135224...  &goetageblatt &1&2&1\\
140697...  &galottom &1&0&0\\
146632...  &1\_FCM &171&26&35\\
135269...  &covid\_watch &0&0&0\\
{...} & ... &...&...&...\\
\bottomrule
\end{tabular}\\[10pt]
\end{table}
\begin{figure}[t]

{\centering \includegraphics[width=0.9\linewidth, trim=1cm 1cm 1cm 1cm, clip]{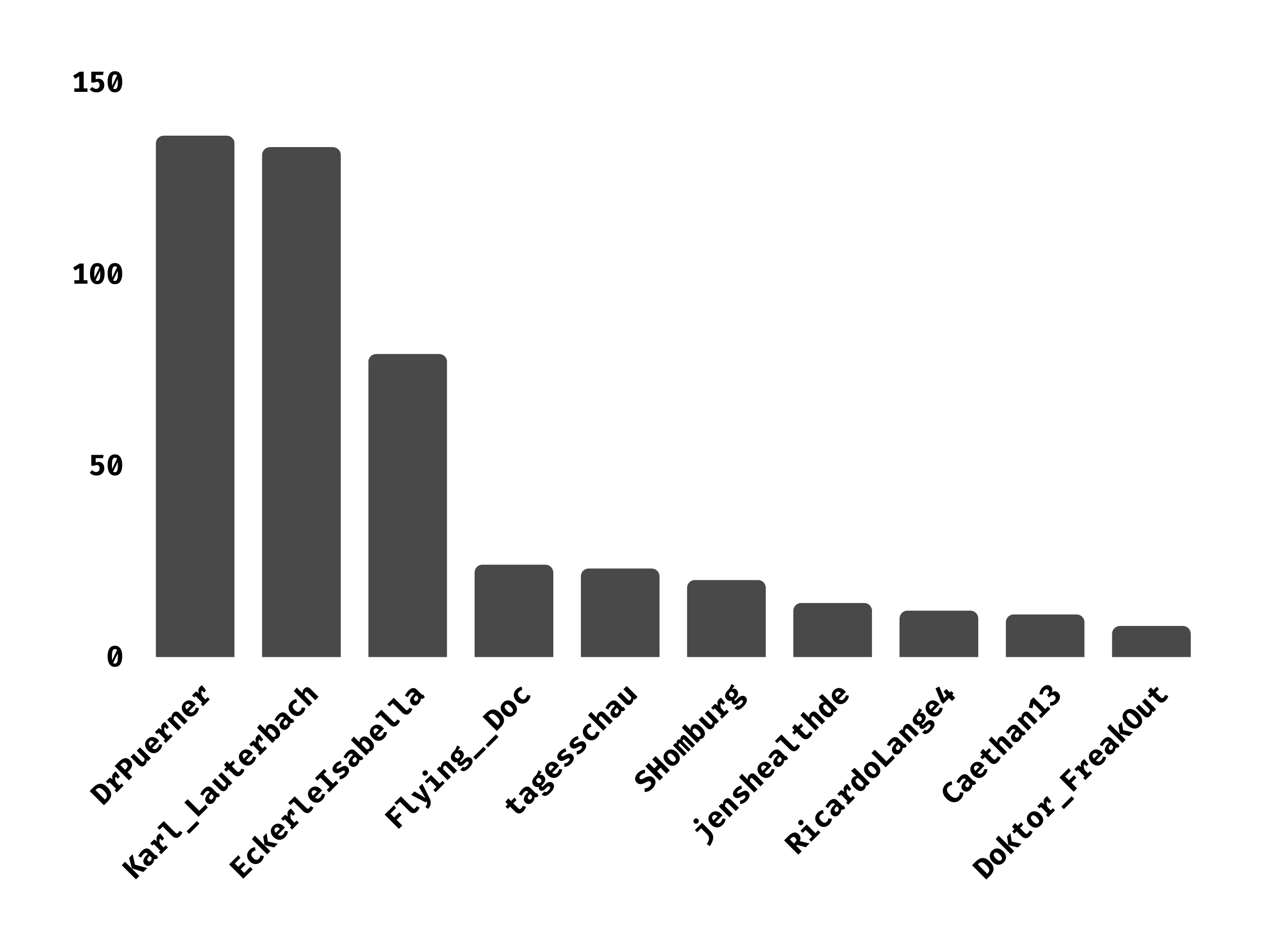} 

}

\caption{Number of tweets authored by the ten most popular users in our sample.}\label{fig:proportion}
\end{figure}
The bar plot in Figure \ref{fig:proportion} depicts the number of tweets authored by the top 10 most popular users in our final sample, displayed by their Twitter usernames. From it, it is apparent how certain actors play a very prominent role in the conversation, with some accounts having authored more than 100 popular tweets in our 1\% sample, meaning that one can expect them to have as much as 100 times more than that overall. This tells us how truly influential elites can be on Twitter, and also indicates that, given the sheer amount of popular tweets by the most prominent accounts, it is quite likely that they will be captured in our 1\% sample.

\begin{figure}

{\centering \includegraphics[width=0.99\linewidth, trim=0cm 7.6cm 0cm 7.6cm, clip]{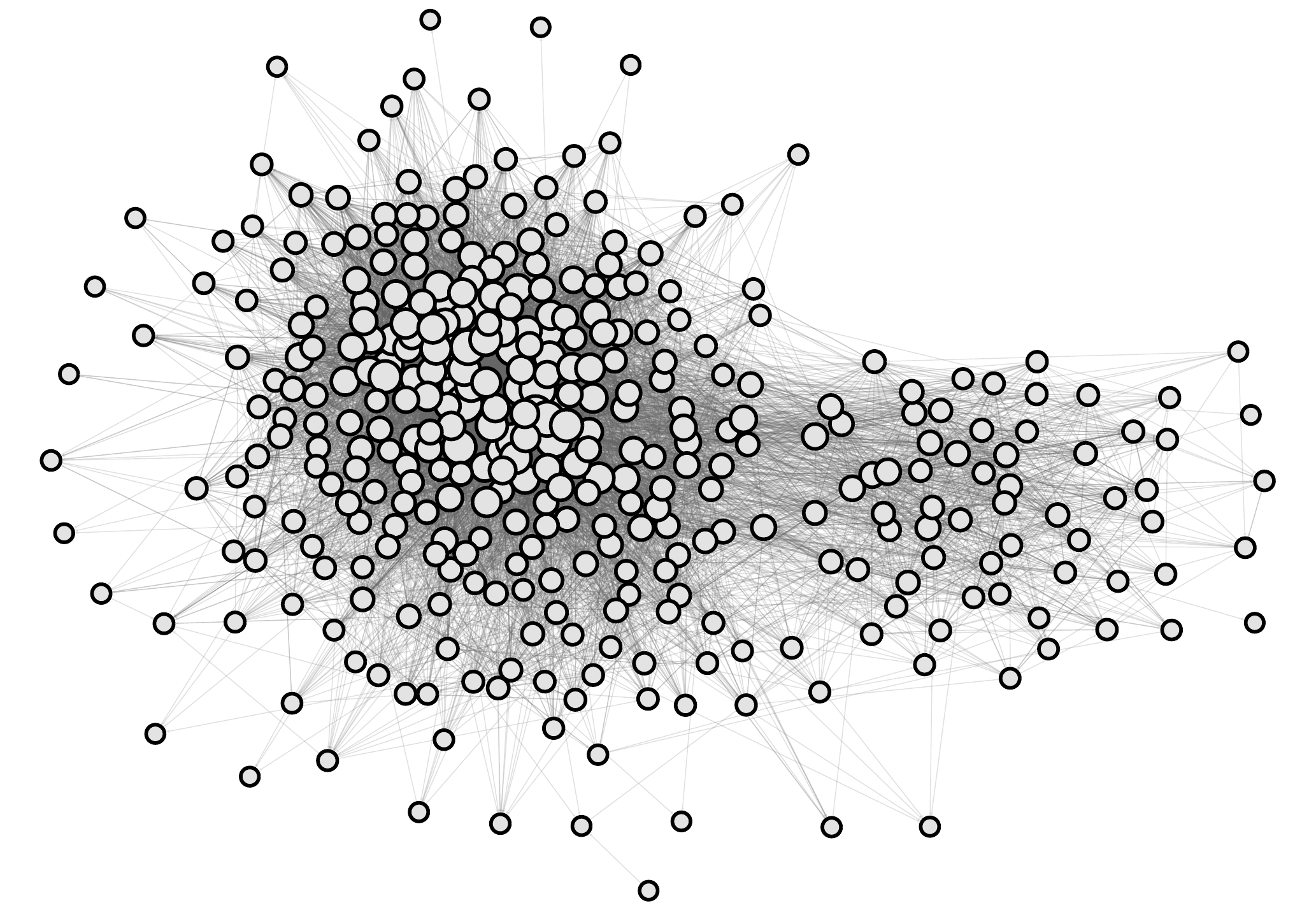} 

}

\caption{Graphical representation of the network of COVID-19 elites on German-speaking Twitter. }\label{fig:network}
\end{figure}

After pinpointing these accounts as COVID-19 elites, we are able to define their following network in a natural way. Specifically, we consider the users as the nodes, and establish that a (directed) edge from actor $i$ to actor $j$ is present if, at the time of the analysis, $i$ follows $j$. After removing the only nine users with no connections, the resulting network is composed of 363 nodes connected by a total of 12182 edges, and is visualized in Figure \ref{fig:network}. From the plot, it is immediately apparent that the network is quite dense: In fact, 9.2\% of all possible edges are observed. Given that the network is composed of users who produced popular tweets about the same topic, the fact that many of them follow each other makes intuitive sense. Moreover, from the graph representation, laid out using a variant of the Yifan Hu force-directed graph drawing algorithm \cite{hu2005}, the network seems to be approximately split into two main groups of different sizes. This already gives a first impression of the two main poles in the network, which will be investigated in more detail in the Results section.

\subsubsection*{Latent space models for social network data}
To model the network data, we make use of the latent cluster random effects model for social
networks \cite{krivitsky2009}.
This model is part of the general family of latent space models, originating from the latent distance model proposed by Hoff et al. \cite{hoff2002}. Latent space network models postulate that each actor has an unobserved position in a $d$-dimensional Euclidean latent social space, and that the probability for two actors to form an edge is inversely related to their distance in the space. This family of models is particularly suitable for social networks, in which mechanisms such as homophily and triadic closure often play a major role \cite{rivera2010}.
Handcock et al. \cite{handcock2007} added the idea of model-based clustering to the original latent distance model, allowing for the actors' positions in the latent space to come from a mixture of normal distributions, where each mixture component represents a cluster. Krivitsky et al. \cite{krivitsky2009} further extend this by adding nodal random effects to control for actor-specific heterogeneity in the propensity to form edges. More precisely, without the inclusion of nodal or edgewise covariates, the model specifies the probability of an edge $y_{ij}$ between nodes $i$ and $j$ through:
\begin{equation}
\begin{split}
 \text{logit}(\mathbb{P}(y_{ij} = 1 | \beta_{0}, \boldsymbol{Z}, \boldsymbol{\delta}, \boldsymbol{\gamma})) = \\ 
\beta_{0} - \lVert \boldsymbol{z}_{i} - \boldsymbol{z}_{j} \rVert + \delta_i + \gamma_j 
\end{split}
\end{equation}
where $\boldsymbol{Z} = (\boldsymbol{z}_1,...,\boldsymbol{z}_n)$ are the latent positions of the nodes in the $d$-dimensional latent space, $\beta_{0}$ is an intercept, and $\boldsymbol{\delta} = (\delta_1,...,\delta_n)$ and $\boldsymbol{\gamma} = (\gamma_1,...,\gamma_n)$ are node-specific sender and receiver effects that account for the individual users' propensity of following or being followed, respectively. Here the latent positions $\boldsymbol{Z}$ are assumed to originate from a finite spherical multivariate mixture of independent normal distributions, and the random effects $\boldsymbol{\delta}$ and $\boldsymbol{\gamma}$ are assumed to be drawn independently from normal distributions with mean $0$ and variances $\sigma^2_{\delta}$ and $\sigma^2_{\gamma}$, respectively. The model is estimated through the \texttt{R} package \texttt{latentnet}, which implements a Bayesian routine based on the use of a Markov chain Monte Carlo algorithm  \cite{krivitsky2008}. It is interesting to note that this model can be viewed as a generalization of the (latent) fitness model for networks \cite{caldarelli2002,bianconi2001}, as the node-specific random effects $\delta_i$ and $\gamma_i$ can be seen as measuring the intrinsic fitness of node $i$ to send and receive ties, while its latent position $\boldsymbol{z}_{i}$ affects its probability of forming ties differently for each (potential) connection.

Homophily and triadic closure are generally prevalent in social media, particularly on Twitter and between popular accounts \cite{lou2013,colleoni2014}. Those mechanisms often lead to the formation of sub-groups of actors based on shared beliefs or other characteristics. 
Identifying such clusters can be helpful in understanding the drivers of polarization and, more in general, grouping behavior.
The general task of identifying assortative, tightly knit groups in networks is a large area of research, known under the umbrella term of ``community detection'' \cite{fortunato2016}. Notable examples of such methods include modularity maximization algorithms \cite{newman2006} and stochastic blockmodels \cite{denicola2022}.
Classical community detection techniques are well suited for finding group structures, but they have the drawback of only returning a discrete partition of the network into clusters, where the connectivity behavior of each actor is fully described by its group label. In other words, two nodes in the same group are considered identical in all aspects. This is generally quite simplistic for social networks, in which cohesive groups often do exist, but where members of each group can also be very different from one another. Within a single group, for example, some nodes might be more ``extreme'' and isolated from all other communities. In contrast, others might be more central to the network, and have many connections to other groups. We expect this to be the case in our network of COVID-19 Twitter elites: While we can assume polarization and grouping behavior to be present, we also expect the social positioning and political beliefs of the actors to be more accurately described through a continuous, multidimensional spectrum rather than with discrete labels. Because of that, we are not only interested in the clear-cut grouping of nodes, but also in uncovering the (continuous) social positioning of the users relative to one another.  
The chosen latent cluster random effects model is, therefore, particularly well suited for our application, as it combines clustering and latent position modeling, thereby enabling us to simultaneously capture polarization and grouping behavior as well as the positioning of the actors relative to each other in the socio-political spectrum.

\section*{Results}
\label{sec:results}

\begin{figure*}[t!]
\setlength{\fboxsep}{0pt}%
\setlength{\fboxrule}{0pt}%
\begin{center}
\includegraphics[scale=0.224, trim=0.2cm 14cm 0.2cm 17cm, clip]{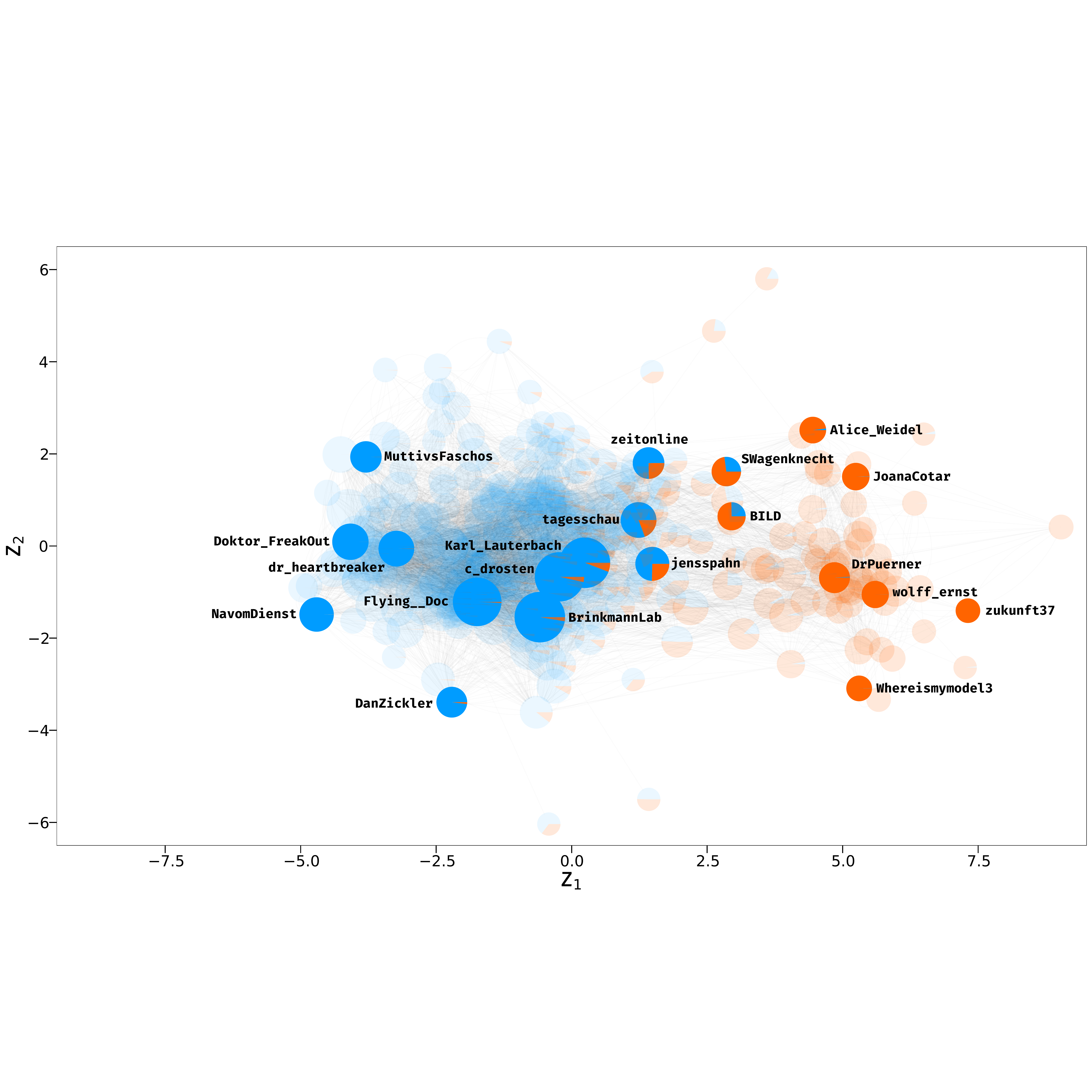}
\end{center}
\caption{Graphical representation of the latent positions of the actors in the network of COVID-19 Twitter elites estimated via the latent cluster random effects model, where the node size for each actor is scaled by its degree. A number of notable users are highlighted. The axes correspond to the two latent dimensions $Z_1$ and $Z_2$, while the estimated posterior probabilities for each user to belong to the ``pro vaccine mandates'' (blue) or ``anti compulsory vaccination'' (orange) cluster are depicted through the node-specific pie charts. Major German media outlets are found between the two communities. 
\label{fig:latentspace}}
\end{figure*}

We fit the latent cluster random effects model to our data, setting both the number of clusters $k$ and the number of dimensions $d$ to 2. The choice of two clusters is backed by the approximated Bayesian Information Criterion for data-driven model selection proposed by Handcock et al. \cite{handcock2007}. Moreover, since much of the literature concerns itself with investigating polarization in the online discussion revolving around the COVID-19 pandemic, and given that polarization suggests the existence of two sub-groups \cite{guerra2013}, setting $k=2$ appears to be the natural choice from a substantive perspective. With regards to the choice of $d$, while dimensionality for latent space network models is generally an open question, setting $d = 2$ is considered to be the standard for applications in which interpretability of the positions is central, as it simplifies the visualization and description of social relationships \cite{sosa2022}. We also experimented with different values of $d$, and observed that using higher dimensionality did not greatly impact the cluster assignments.

The results of the model fitting are visualized in Figure \ref{fig:latentspace}. The axes correspond to the two latent dimensions $Z_1$ and $Z_2$, respectively, and the nodes' colors indicate the estimated community memberships. More specifically, the node-specific pie charts represent the posterior probabilities for each user to belong to the one or the other cluster. Node sizes are scaled by each actor's total degree within the network. Note that, as defined by the model, two nodes that are closer to one another have a higher probability of forming an edge, i.e.\ of following each other. Also note that estimates of the node-specific random effects $\boldsymbol{\gamma}$ and $\boldsymbol{\delta}$, incorporating information on how active specific nodes are with respect to following or being followed, are made available in the supplementary materials. At first glance, we see that the two communities are distributed along the horizontal axis $Z_1$, with the more numerous blue community occupying the left and center parts of the figure, and the orange one being located towards the right-hand side. Moreover, from the posterior membership probabilities we can see that group memberships are fairly clear for most nodes. Nonetheless, significant uncertainty can be observed for a non-negligible proportion of the actors, which lie in between the two clear communities in the space.  

As our task is of unsupervised nature, we do not have a set-in-stone ``ground truth'' with which to compare the model-based labeling and the estimated positions of the actors. To interpret the results, we therefore need to dig into the data and consider the emerging patterns. As the network is limited in size, and thanks to the naturally high propensity of elite users to voice their opinions, it is relatively straightforward to identify some of the more prominent actors and gauge their views on pandemic-related governmental interventions based on public information. 
Through this process, we can appreciate how the latent position of each actor in the network is strongly associated with their public stances on government mandates. More specifically, despite substantial within-cluster heterogeneity in stances (and their intensity) on several issues, users in the blue community hold views that can be summarized as ``generally for'' interventions and vaccine mandates. The opposite is true for actors in the orange community, which can be described as ``generally against'' such measures. Moreover, the positioning of nodes within communities is also informative on the actors' beliefs, capturing the within-cluster heterogeneity mentioned. Specifically, more central (external) positions in the overall latent space are associated with more moderate (extreme) stances. To showcase these patterns, we highlighted and labeled some notable users in Figure \ref{fig:latentspace}, where each user is indicated with their Twitter username. The very center of the space is occupied by the most popular actors, most of whom, despite having connections to both groups thanks to their ``elite among elites'' status, reside firmly in the blue camp: A prime example is \texttt{Karl\_Lauterbach}, an exponent of the Social Democratic Party (SPD) who, at the time of writing, has been serving as the health minister of Germany since December 8, 2021. He is known to be a strong proponent of vaccination and mandatory vaccination for all \cite{lauterbach}. Two other notable members of this group are Christian Drosten (\texttt{c\_drosten}), a prominent virologist who has been described by major media outlets as ``the country's real face of the coronavirus crisis'' and ``the nation's corona-explainer-in-chief'' \cite{drosten}, and Melanie Brinkmann (\texttt{BrinkmannLab}), another well-known virologist who was among the proponents of the No-COVID strategy \cite{brinkmann}.
Moving a bit further left in the space, another very popular user in the network is \texttt{Flying\_\_Doc}, a medical doctor who has been outspoken in his support for policy proposals such as a vaccine mandate for all adults, and allowing access to events only to people who are both fully vaccinated and tested (``1G+'' in the German political jargon).
Looking even more toward the left on the $Z_1$ dimension, we encounter positions that are increasingly more in the direction of decisive government interventions. Examples of this are  \texttt{dr\_heartbreaker}, a medical professional who has expressed his support for hard lockdowns and the aforementioned No-COVID strategy, and \texttt{NavomDienst} and \texttt{Doktor\_Freakout}, two anonymous medical doctors who also vehemently voiced their dissent for what they deemed to be bland policy making, and vouched their support for stronger restrictions.
To conclude our outlook on the blue community, we also labeled two more peripheric, less Twitter-popular nodes. On the bottom-left of the plot we find \texttt{DanZickler}, an intensive care doctor who also expressed his support for more decisive action by the government, while on the top left we find \texttt{MuttivsFaschos}, who tweeted at the hashtags \texttt{\#ZeroCovid} and \texttt{\#harterLockdownJetzt} (``harder lockdown now'').
All in all, our analysis highlights how users categorized in the blue group generally tend to openly support governmental efforts to contain the pandemic, and that the estimated dimension $Z_1$ is associated with the intensity of the actors' voiced stances on policy. 

We now shift our focus to the orange community, composed of actors who have, on average, significantly fewer followers within this elite network, and tend to more or less strongly oppose pandemic-related government mandates.
We start our overview with \texttt{DrPuerner}, the user with the highest number of popular tweets in our dataset. A medical doctor, Puerner rose to prominence during the pandemic for his stark criticism of COVID measures and opposition to government mandates. While not downplaying the dangers posed by COVID-19, he attracted following and praise from conspiracy theorists and the populist right-wing party ``Alternative for Germany'' (``AfD''), notorious for its anti-system beliefs \cite{puerner}. Closer to \texttt{DrPuerner} in the latent space we can also find \texttt{wolff\_ernst}, a self-described journalist and writer, who has openly associated himself with COVID-related and general conspiracy theories \cite{wolff}. 
We also labeled two more peripheral nodes in this cluster, namely users \texttt{zukunft37} and \texttt{Whereismymodel3}, anonymous accounts who openly voice their vaccine skepticism and opposition to government mandates.
Two elected members of the aforementioned AfD, namely \texttt{Alice\_Weidel}, who has been the leader of the party in the Bundestag (German Federal Parliament) since October 2017, and  \texttt{JoanaCotar}, another member of the Bundestag who was part of AfD for the whole studied period and until late 2022, are also part of the orange community. Unsurprisingly, the two are close in the latent space, reflecting their similar policy stances. Perhaps more surprisingly, their estimated latent positions are not far from that of Sahra Wagenkecht (\texttt{SWagenknecht}), member of the Bundestag for ``The Left'' (``Die Linke'') since 2009, and former parliamentary leader of that same party. Despite being on the other end of the political spectrum, she also opposes general vaccination mandates \cite{wagenknecht}. She is located more towards the middle of the plot, and has substantial uncertainty in her community membership, with a posterior probability of approx. 75\% to belong to the orange community. Another actor whose community membership is uncertain is Christian Democratic Union (CDU) politician Jens Spahn, who served as health minister for most of the analyzed period, i.e.\ until December 8, 2021. (\texttt{jensspahn}). He is not far in the space from his successor Karl Lauterbach, but lies a bit more on the right: He is classified in the blue community, but has a posterior probability of approximately 25\% to belong to the orange one. This is in line with the fact that, while he is a proponent of widespread vaccination, he is opposed to the idea of compulsory vaccination for all \cite{spahn}. To conclude our overview of the space, we highlight some other notable accounts located in between the two clusters, namely those belonging to prominent news outlets. Given that we expect them to have a diverse following due to their authority status, their central positioning makes intuitive sense. But even between media outlets, the model is able to draw a distinction: \texttt{zeitonline} and  \texttt{tagesschau}, generally reputable news sources, are closer to the center of the space, and, although with substantial uncertainty, labeled as blue. On the other hand, \texttt{BILD}, the most prominent German boulevard newspaper, is located more towards the right, and has a higher probability of belonging to the orange group.

\section*{Discussion}

In this paper, we identified and modeled the network of users leading the conversation revolving around the COVID-19 pandemic on Twitter. More specifically, we made use of the latent cluster random effects model to map these elite users into a two-dimensional Euclidean social space, in which users that are closer to each other have a higher likelihood to connect, i.e.\ to follow each other.  
The results suggest the emergence of a natural partition of the network into two dense macro-communities, which are only loosely connected with their opposing counterparts. 
By focusing on a number of notable users, such as politicians, activists, and news outlets, we show how those two communities
can be interpreted as ``generally pro'' and ``generally against'' public interventions and vaccine mandates. This finding corroborates recent research demonstrating the polarized nature of pandemic-related online discourse, especially concerning vaccination \cite{jiang2021,reiter2022polarization,steelfisher2021,cowan2021,monsted2022}. But a deeper inspection of the latent space further reveals that users within communities are only partially homogeneous in their stances. To the contrary, the model is able to uncover a nuanced, continuous spectrum of pandemic-related beliefs and policy positions, ranging from demanding radical containment measures all the way to vaccine skepticism and COVID-denying conspiracy theories, covering everything in between those two extremes. In this context, neutral actors, such as mainstream news outlets, are positioned between the two clusters, which makes intuitive sense given their authority status. From the latent positions of users whose political inclination is known, we can also appreciate how attitudes toward governmental interventions tend to follow political inclination, with left and right-wing respectively corresponding to more favorable or unfavorable positions towards restrictions and vaccine mandates. This finding echoes recent research showing how ideology can shape trust in scientists and attitudes towards vaccines \cite{featherstone2019,kossowska2021}. The importance of vaccination as a sub-theme within the pandemic-related discussion is corroborated by the fact that ``vaccine'' is one of the words appearing more often in the data (while not being used as a filter), as shown in the supplementary material (Figure \ref{fig:wordcount}). 

A particular feature of the employed methodology is the ability to combine ``classical'' community detection, which alone would be insufficient to gain a proper understanding of the network at hand, with more refined, continuous latent space modeling. 
This allows to map the underlying latent social space with the necessary nuance while simultaneously returning a partition of the network into sub-groups, which can be useful for understanding the network at a coarser resolution, or for classification purposes. The modeling results thus allow us to obtain a clearer picture of the network as a whole, and can be used for garnering insight on single (politically unaffiliated) users.

 We note that the studied network is fairly small as a result of the relatively restrictive popularity threshold we chose for defining a popular tweet: It would thus be possible to decrease the threshold to obtain a larger network. We also note, however, that using a lower value somehow ``loosens'' the definition of an elite, as users that are less popular on average would make it into the network. Experimenting with the threshold, we also observed that using different values almost only impacts the size of the network's periphery, and does not change the overall picture. Results of alternative analyses with different threshold values and inclusion criteria are provided in the supplementary material (Figures \ref{fig:robustness} and \ref{fig:confusion}), and corroborate the robustness of our findings. Moreover, a stricter definition of elites incidentally makes the network size more manageable, which is relevant given that model estimation, as it is currently implemented in the \texttt{R} package \texttt{latentnet}, only scales well up to a few thousand nodes. Nonetheless, while latent space models do pose serious computational challenges, different approaches to estimate them for larger networks have been proposed  \cite{raftery2012, yin2013}. 
We also note that, as we here only model the behavior of elites, we cannot \textit{a priori} assume our results to be valid for the overall discussion. While, given the well-documented strong influence of popular users in the conversation, it is reasonable to believe that many of the results could extend to the general Twitter population, further research would be needed to confirm this.  
Furthermore, there may be different patterns in how elite and non-elite actors follow other users. For example, whereas non-elites are likely to use their follows primarily instrumentally, i.e., to see tweets they are interested in on their timeline, elites could also use theirs for signaling, i.e., to publicly show support or endorsement towards other users, and may thus curate their follows more carefully. Similarly, highly active elites could be more likely than non-elites to enter conflicts with each other and block opposing elites. On the one hand, these strategic follows (or non-follows) are indeed relevant to our analysis, as they give information on the potential factions at play in the network, and aid us in identifying them. On the other hand, as a result of these mechanisms, polarization in the elite network may be higher than in the complete one. The latter consideration strengthens the notion that polarization, although undoubtedly present to some extent, is not the only determining factor in network formation, and that the different groups exist on a continuous spectrum rather than being completely isolated from one another.

We further emphasize that our approach is purely unsupervised and completely based on network structure, without including any element of natural language processing. In other words, this means that the two groups emerge only from using information on who follows whom. In this sense, we could have simply labeled the two clusters as ``blue'' and ``orange'', or ``left'' and ``right''. The description of the communities with respect to their attitudes towards vaccination, and, more in general, pandemic management, was done after the modeling, to shed some additional light on the data-driven cluster selection, and alternative characterizations would also be viable. While it would certainly be possible to make use of the tweets' text content to obtain further insight into the users, we here explicitly chose to focus solely on the network component, thus demonstrating how tightly the users' personal networks are intertwined with their beliefs. Indeed, given that the latent positions of the actors are estimated by the model solely using their follows and followers within the network, it is quite remarkable how consistently actors neighboring each other in the estimated latent space are also near in their stances on COVID-19 and its management, and how closely the space is able to track the belief spectrum. The echo chamber effect is well documented in the literature: Users tend to follow those who share similar ideas, and are thus rarely exposed to contrasting views. This, in turn, leads the users' beliefs to become self-reinforcing \cite{cinelli2021, nguyen2020}. 
However, our analysis demonstrates how this behavior is not only prevalent at the extremes of the socio-political spectrum, but also towards the center of the belief space. On the one hand, the phenomenon implies that users with radical ideas will tend to follow people with similarly extreme beliefs, leading to further polarization; On the other hand, it also means that users following more moderate voices will also tend to gravitate towards more nuanced views. Central actors, which have the ability to act as a bridge between the two communities, are thus uniquely positioned to mitigate the polarization loop. 

The fact that following behavior is so closely related to beliefs and attitudes paves the way for latent space models as powerful tools for drawing maps of social media landscapes, which can, in turn, be used to increase our understanding of the underlying social and behavioral structures.  
Indeed, while we here applied the methodology to map the discussion revolving around COVID-19, it is possible to perform similar types of analysis on other topics of public relevance.
Given its explanatory and predictive power, we believe latent space modeling of elite social media networks to have the potential for improving our general understanding of the online landscape, ultimately aiding policymakers in making more informed decisions in their quests against polarization and misinformation worldwide.

\section*{Data Availability}
The network data used in this paper and the code to reproduce the analysis are publicly available on our GitHub repository, at \href{https://github.com/gdenicola/latent-space-covid-twitter-elites}{https://github.com/gdenicola/latent-space-covid-twitter-elites}.

\section*{Acknowledgements}
This research was partially funded by the Elite Network of Bavaria (ESG Data Science). 



\printbibliography

\end{refsection}

\newpage

\renewcommand{\thefootnote}{\fnsymbol{footnote}} 
\begin{refsection}
\onecolumn

\begin{center}
	{\LARGE{}{{Supplementary Material}} \\
 \medskip
 \LARGE{\textbf{COVID-19 and social media: Beyond polarization}}}  \\ 
	Giacomo De Nicola$^{a,}$\footnote[1]{Corresponding author:  giacomo.denicola@stat.uni-muenchen.de}, Victor H. Tuekam Mambou$^{b,a}$, Göran Kauermann$^{a}$\hspace{.2cm}\\
	\vspace{0.3cm}
	Department of Statistics, LMU Munich$^a$ \\
ifo Institute -- Leibniz Institute for Economic Research at the University of Munich, Germany$^b$\hspace{.2cm}
\end{center}

\setcounter{section}{0}
\setcounter{page}{1}

\makeatletter 
\setcounter{figure}{0}
\setcounter{table}{0}
\renewcommand{\thefigure}{S\@arabic\c@figure}
\renewcommand{\thetable}{S\@arabic\c@table}
\renewcommand{\thesection}{S\@arabic\c@section}

\makeatother

\setcounter{equation}{0}

The following contains technical details and supplementary information to the manuscript. The full data and code to reproduce the analysis and to aid the reader in applying the presented models to their own data can be found in our {Github repository}, available at \href{https://github.com/gdenicola/latent-space-covid-twitter-elites}{\texttt{https://github.com/gdenicola/latent-space-covid-twitter-elites}}.

\section{Defining the set COVID-19 Twitter elites}
Social media elites can be broadly understood as ``users with the ability to influence'' \cite{dubois2014}.
The concept is further widely discussed in the literature \cite{papacharissi2012,jackson2016,gallagher2021}. 
While there is no unique, precise definition of it, the idea can further be narrowed down: The term ``social media elite'' typically refers to a group of highly influential and popular users with extensive reach on the platform and who have a significant impact on conversations, trends, and narratives circulating online. These users often include celebrities, politicians, journalists, thought leaders, and influencers who have a large and engaged audience and are frequently retweeted, quoted, and mentioned by others. This characterization, while informative, is quite broad, and by itself does not indicate a unique way of identifying elites in practice. 
Operational definitions for empirical applications are often based on engagement metrics, such as the number of followers of each user, as well as tweet likes, retweets, quotes, and replies. Choices within these metrics are virtually endless and depend on both the data at hand and the research question that is tackled. In our case, the focus lies on analyzing the behavior of users who actively participate in the Twitter conversation on COVID-19, and that have the ability to influence it, i.e.\ reach many other users. To do so, we made use of a dataset containing a 1\% sample of all tweets containing keywords relating to the pandemic in 2021. Based on that, we opted to identify COVID-19 Twitter elites as accounts who authored popular tweets on the topic, i.e.\ with a total popularity score (given by the sum of likes, retweets, and replies) over a certain threshold, which in our case was set to 2000. Different solutions, such as using the total number of followers, or the total popularity score across all tweets, would also be viable, and were considered. We eventually ended up with our criterion for several reasons:
\begin{itemize}
    \item[(a)] Firstly, using the total number of followers would tend to include general celebrities, such as e.g.\ famous actors or pop stars, who only sporadically tweet about the pandemic, possibly with little engagement. Conversely, this criterion would tend to exclude people who are not generally famous, but play an active role in the COVID conversation, and are thus able to reach many users with some of their tweets. While, of course, tweet engagement for a certain user is highly correlated with their number of followers, our focus lies more on the COVID-related, tweet-specific engagement. We, therefore, opted for using direct measures of tweet engagement instead of following counts in our operational definition. 
    \item[(b)]Secondly, we decided against using the total popularity of all tweets by a given user over the analyzed period to identify elites because we are interested in accounts that can produce tweets reaching many people. In contrast, using total engagement across all tweets (without any cutoff for the popularity of single tweets) would include accounts that may garner many total shares because they produce an abundant amount of low-quality posts that each happen to get a little amplification. Examples of these types of accounts may be bots or spam accounts, as discussed by Gallagher et al. \cite{gallagher2021}. We, therefore, only include users who authored tweets that can be considered impactful on their own. Nonetheless, we thought it would be useful to also carry out a secondary analysis using this alternative criterion based on total popularity, as an additional robustness check for our results. The results of the latter analysis are included in Section \ref{sec:robustness}, and show that our core results are robust to a criterion shift in this direction. 

    \end{itemize}

We finally note that a way to further strengthen our criterion towards ``sustained'' as opposed to ``episodic'' influence would be to require users to have authored at least two popular tweets included in the sample. This definition lowers the risk of single users ending up in the network by pure chance, i.e. by only having produced very few popular tweets without being a ``true'' elite. This, of course, comes at the cost of randomly excluding some true elites who may have been only selected in our 1\% sample a single time despite having many popular tweets. Given that a 1\% sample is already likely to exclude important users, we decided that it was enough to have a single popular tweet to be included. Nonetheless, we additionally ran the analysis using this stricter criterion for comparison purposes, and included the results in the following section.

\section{Robustness checks}
\label{sec:robustness}

In this section, we repeat the analysis performed in the main body of the paper using different criteria for users to be included in the network of COVID-19 Twitter elites.
This is meant to showcase the robustness of our results to variations in the choice of popularity threshold as well as structural changes in the inclusion criterion itself. 
As a reminder, in the main analysis a user was included in the elite network if they authored at least one tweet with popularity score $\ge 2000$, where the popularity score is given by the sum of likes, replies and retweets (including mentions). We here performed four additional analyses, each with a different inclusion criterion. These include:

\begin{enumerate}[label=(\alph*)]
    \item Increasing the popularity threshold from 2000 to 4000, i.e. a user is included in the network if they authored at least one tweet with a total popularity score of 4000;
    \item Lowering the popularity threshold from 2000 to 1000, i.e. a user is included in the network if they authored at least one tweet with a total popularity score of 1000;
    \item Strengthening the definition of elite by requiring at least two popular tweets (i.e. with popularity $\geq 2000$) to be included in the network (instead of at least one tweet); 
    \item Changing the definition of elite by allowing users who obtained a total popularity score $\geq 5000$ across all tweets (instead of in a single one) into the network.
\end{enumerate}

\begin{figure}
     \begin{subfigure}[b]{0.53\textwidth}
         \includegraphics[width=\textwidth]{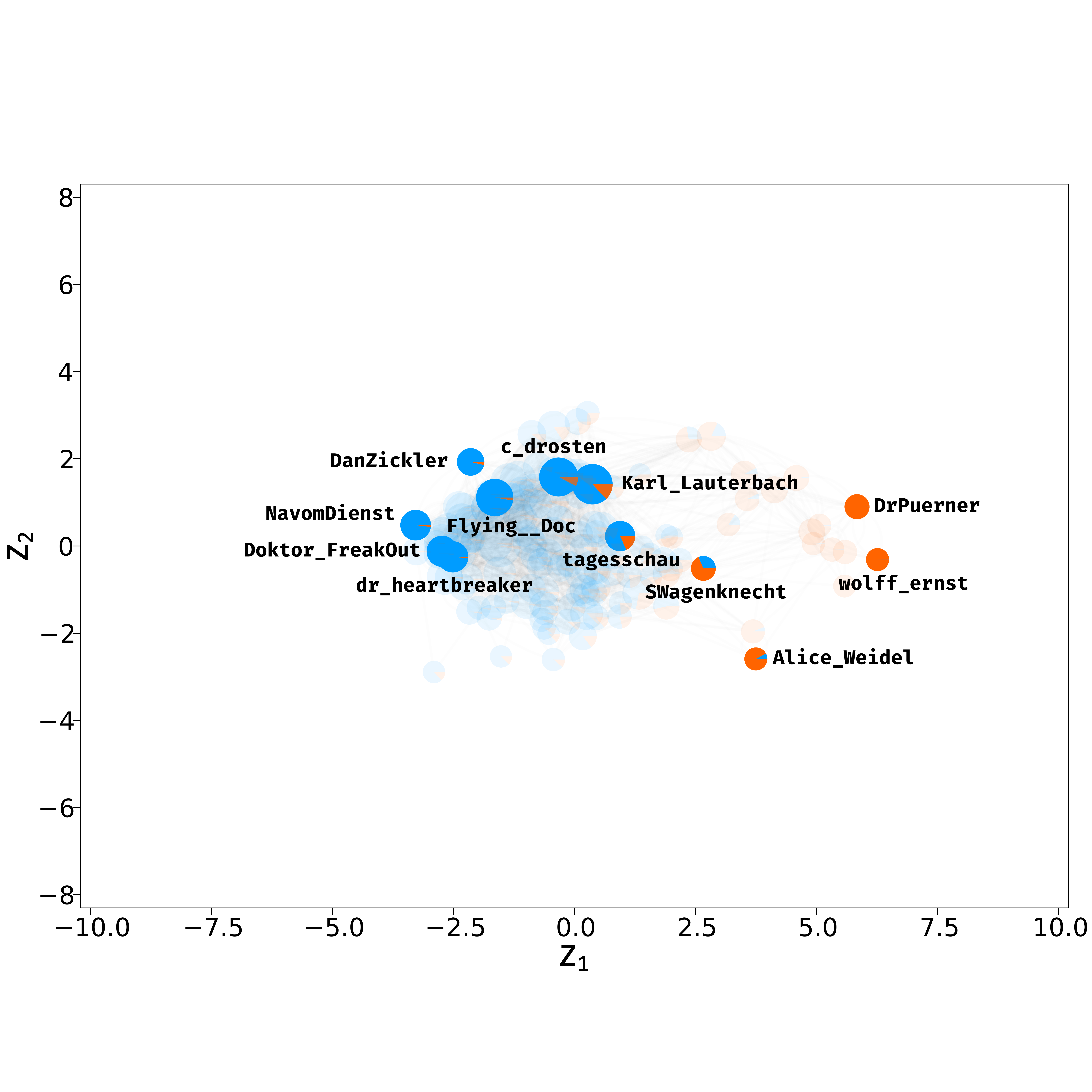}
         \caption{Latent positions of users who authored at least one \\ tweet with popularity score $\ge 4000$ (137 users).}
         \label{fig:smaller}
     \end{subfigure}
     \hfill
     \begin{subfigure}[b]{0.53\textwidth}
         \includegraphics[width=\textwidth]{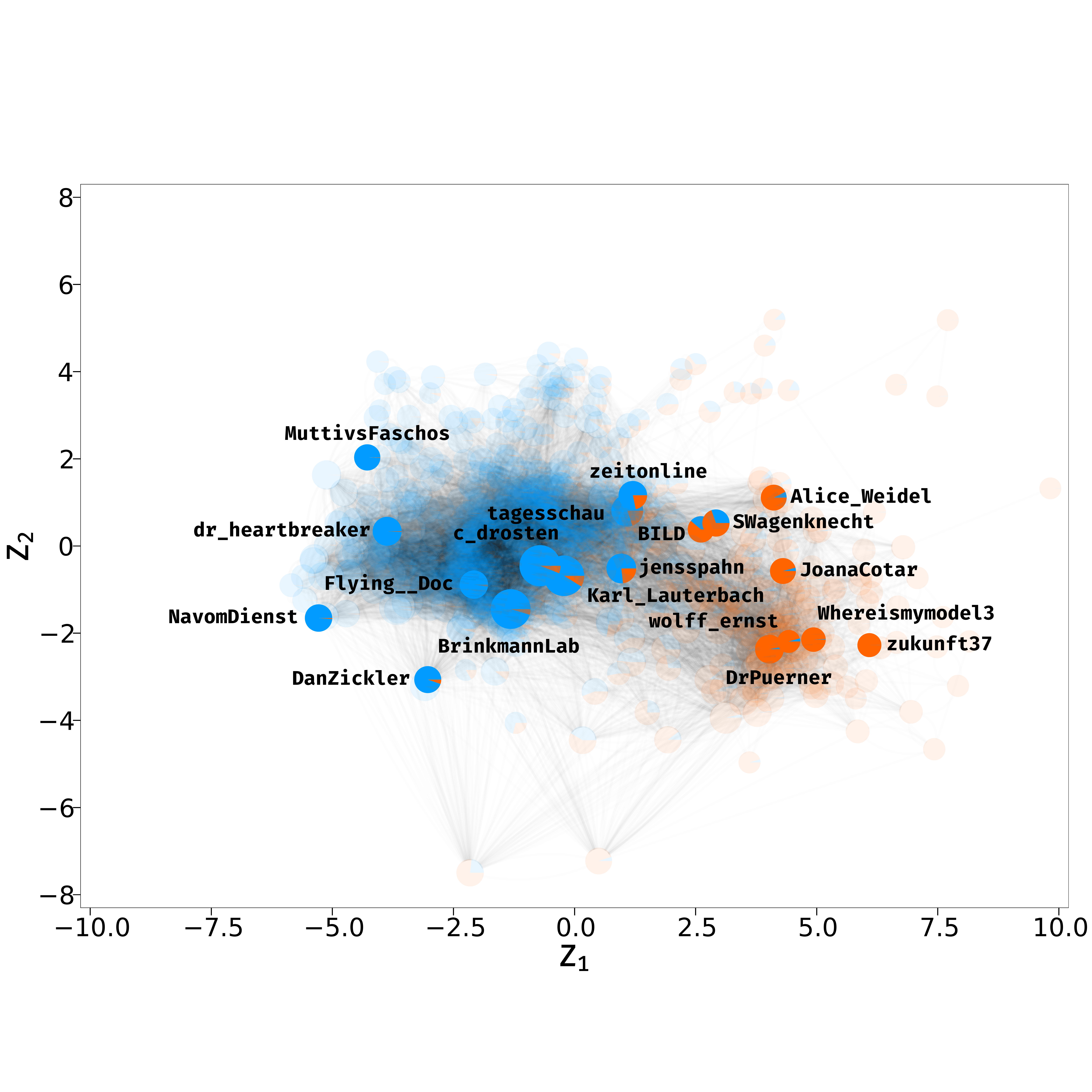}
         \caption{Latent positions of users who authored at least one \\ tweet with popularity score $\ge 1000$ (730 users).}
         \label{fig:larger}
     \end{subfigure}
     \hfill
     \begin{subfigure}[b]{0.53\textwidth}
         \includegraphics[width=\textwidth]{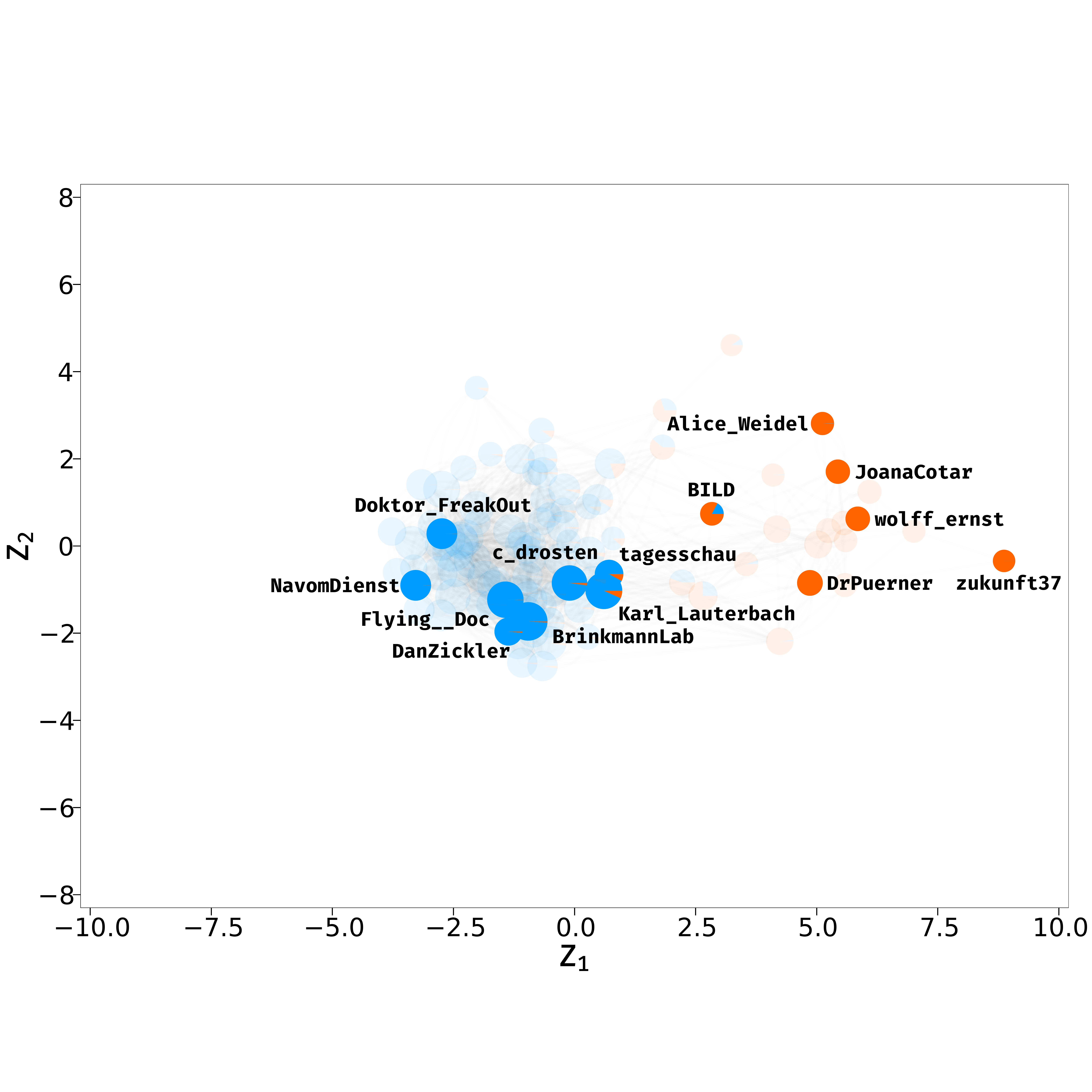}
         \caption{Latent positions of users who authored at least two \\ tweets with popularity score $\ge 2000$ (99 users).}
         \label{fig:atleast2}
     \end{subfigure}
     \hfill
     \begin{subfigure}[b]{0.53\textwidth}
         \includegraphics[width=\textwidth]{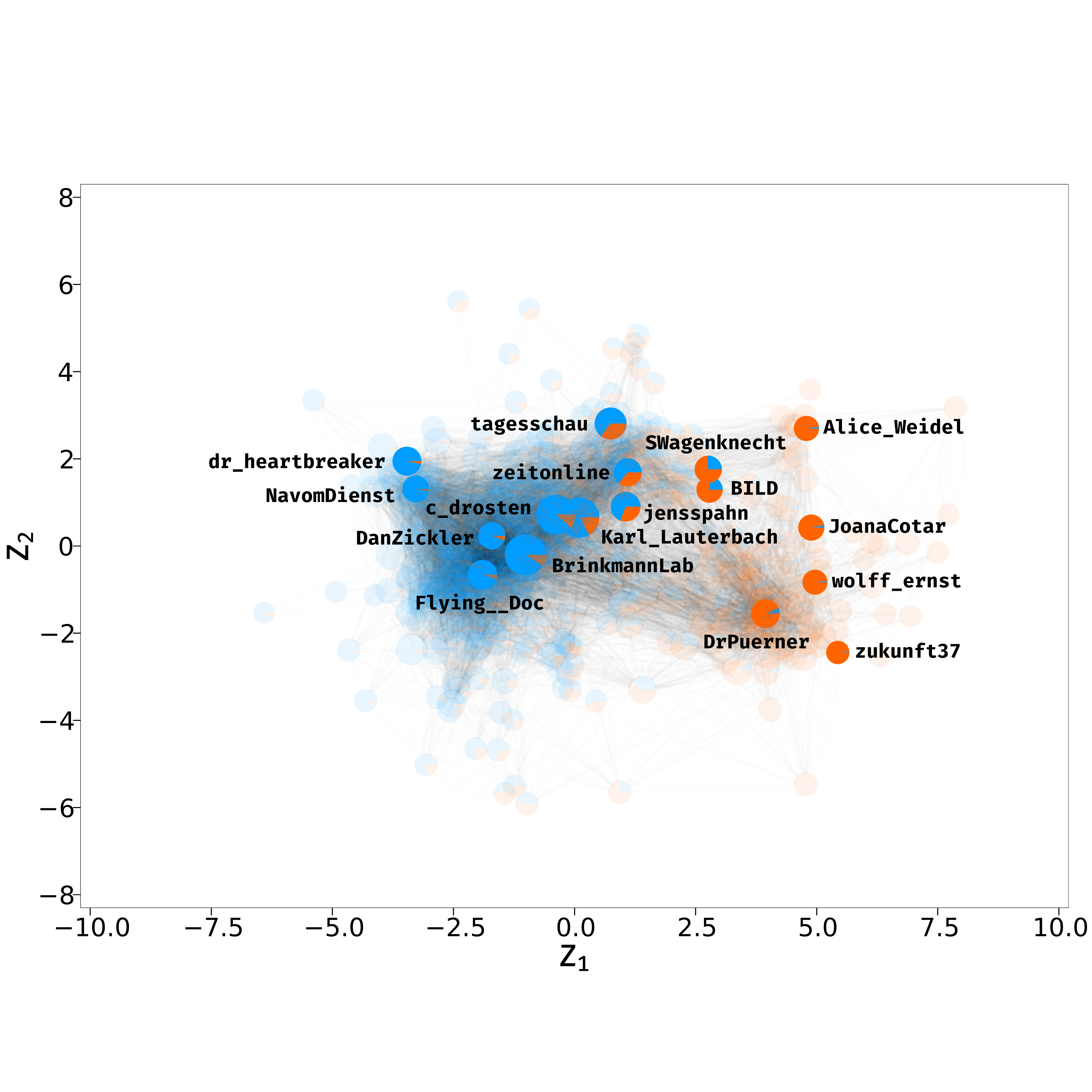}
         \caption{Latent positions of users with a cumulative popularity \\ score across all tweets $\ge 5000$ (516 users).}
         \label{fig:total-pop}
     \end{subfigure}
\caption{Latent positions of the actors in the network of COVID-19 Twitter elites estimated via the latent cluster random effects model using different popularity criteria. A number of notable users are highlighted.} 
    \label{fig:robustness}
\end{figure}

All these different inclusion criterias result in either smaller or bigger elite sets by allowing more or less users into the network. After obtaining the networks, we fit the latent cluster random effects model for each of them, and plotted the results in Figure \ref{fig:robustness} to provide a visual comparison of the positions of the different nodes. From the figure we can appreciate how the size of the network varies significantly with the criterion, but also that, while there is some minor nodal movement, the positions of the labeled actors are quite similar across the different networks. Even in Figure \ref{fig:total-pop}, i.e.\ the one depicting the network with total popularity instead of single-tweet popularity as its inclusion criterion, the positioning of the labeled nodes does not seem to exhibit major differences with respect to the original analysis. 

\begin{figure}
     \begin{subfigure}[b]{0.49\textwidth}
         \includegraphics[width=\textwidth]{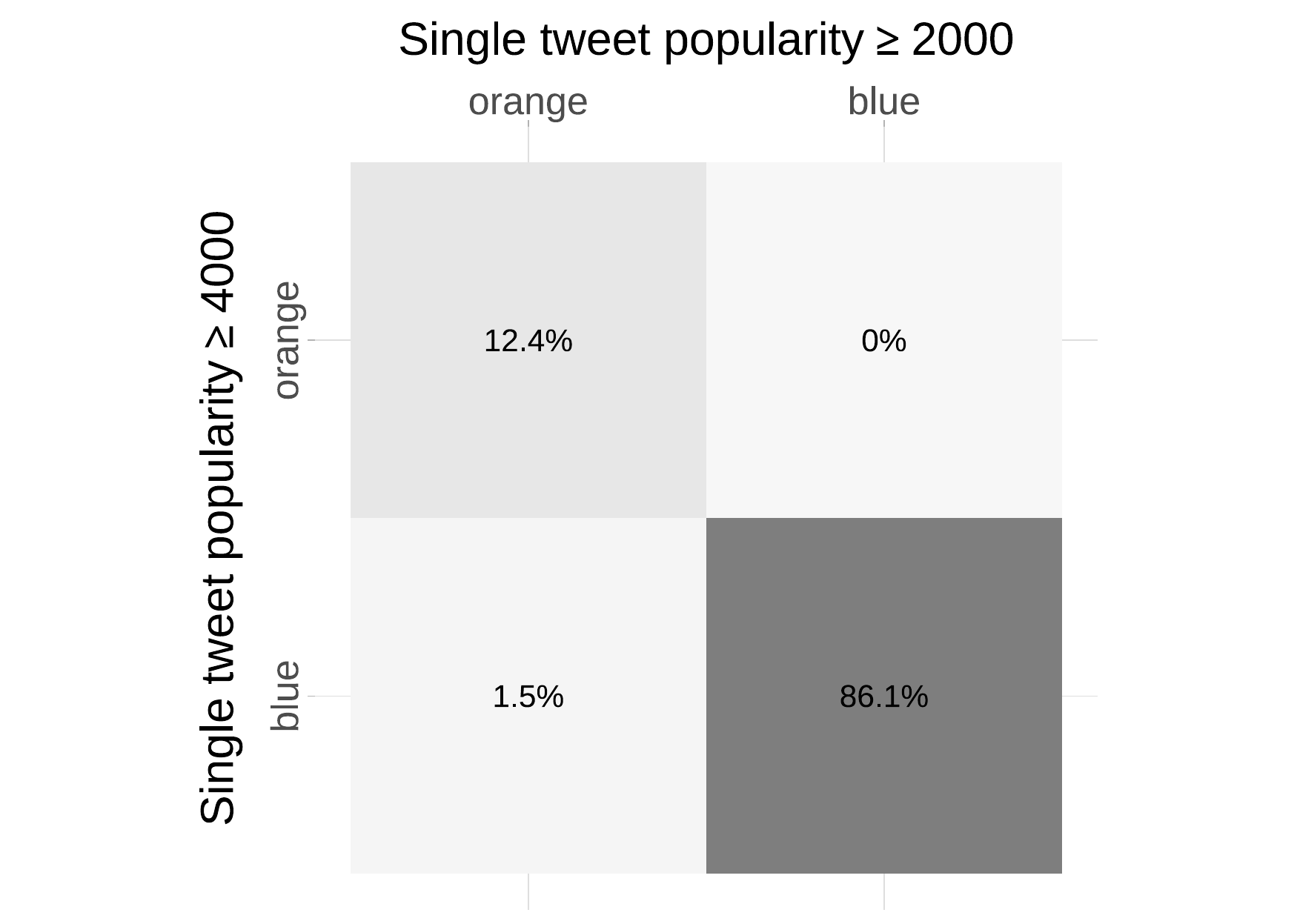}
         \caption{Confusion matrix for classifications in 
 Figure \ref{fig:smaller} \\ versus main analysis (Figure \ref{fig:latentspace}).}
         \label{fig:confusion-smaller}
     \end{subfigure}
     \hfill
     \begin{subfigure}[b]{0.49\textwidth}
         \includegraphics[width=\textwidth]{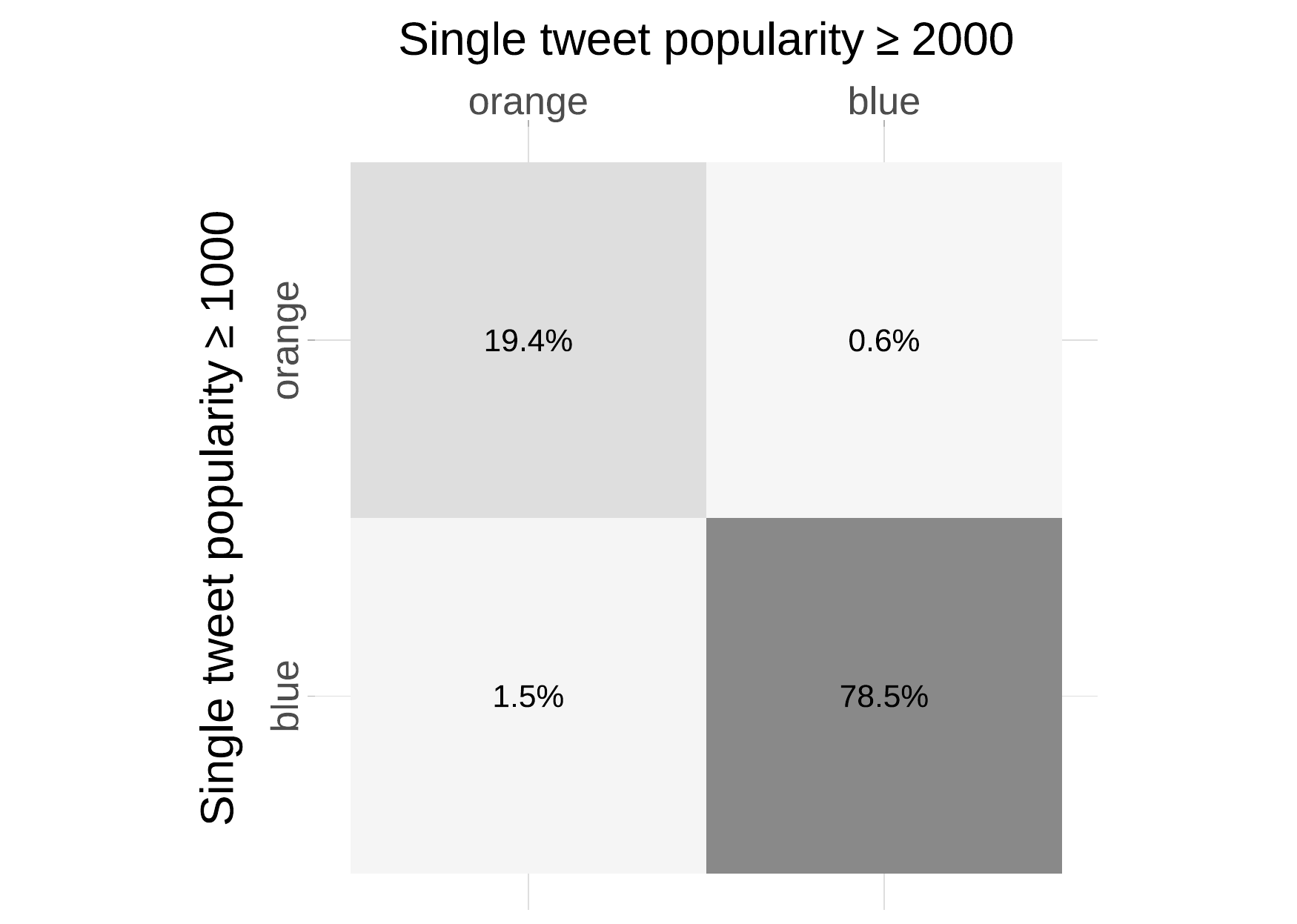}
         \caption{Confusion matrix for classifications in 
 Figure \ref{fig:larger} \\ versus main analysis (Figure \ref{fig:latentspace}).}
         \label{fig:confusion-larger}
     \end{subfigure}
     \hfill
     \begin{subfigure}[b]{0.49\textwidth}
         \includegraphics[width=\textwidth]{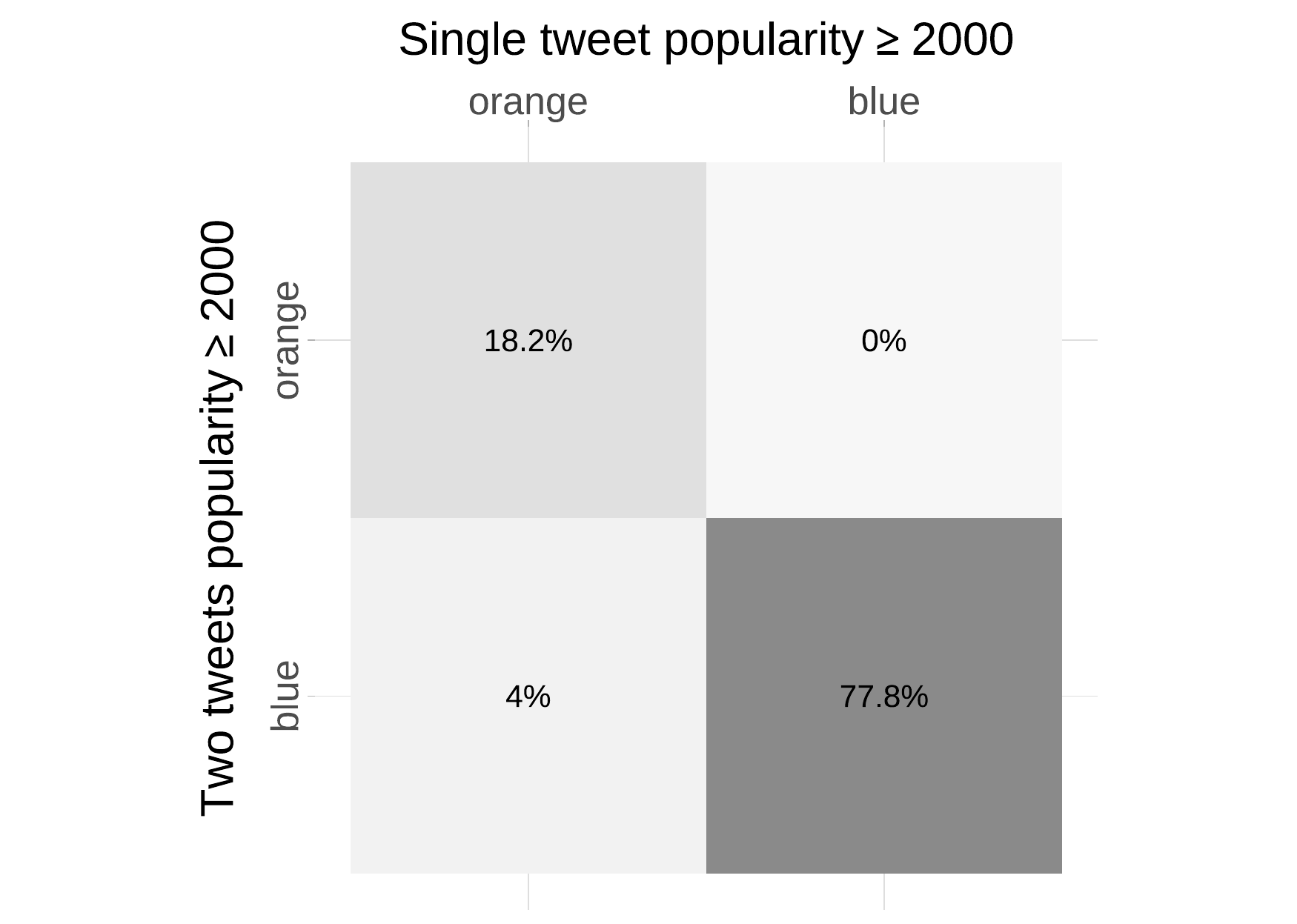}
         \caption{Confusion matrix for classifications in 
 Figure \ref{fig:atleast2} \\ versus main analysis (Figure \ref{fig:latentspace}).}
         \label{fig:confusion-atleast2}
     \end{subfigure}
     \hfill
     \begin{subfigure}[b]{0.49\textwidth}
         \includegraphics[width=\textwidth]{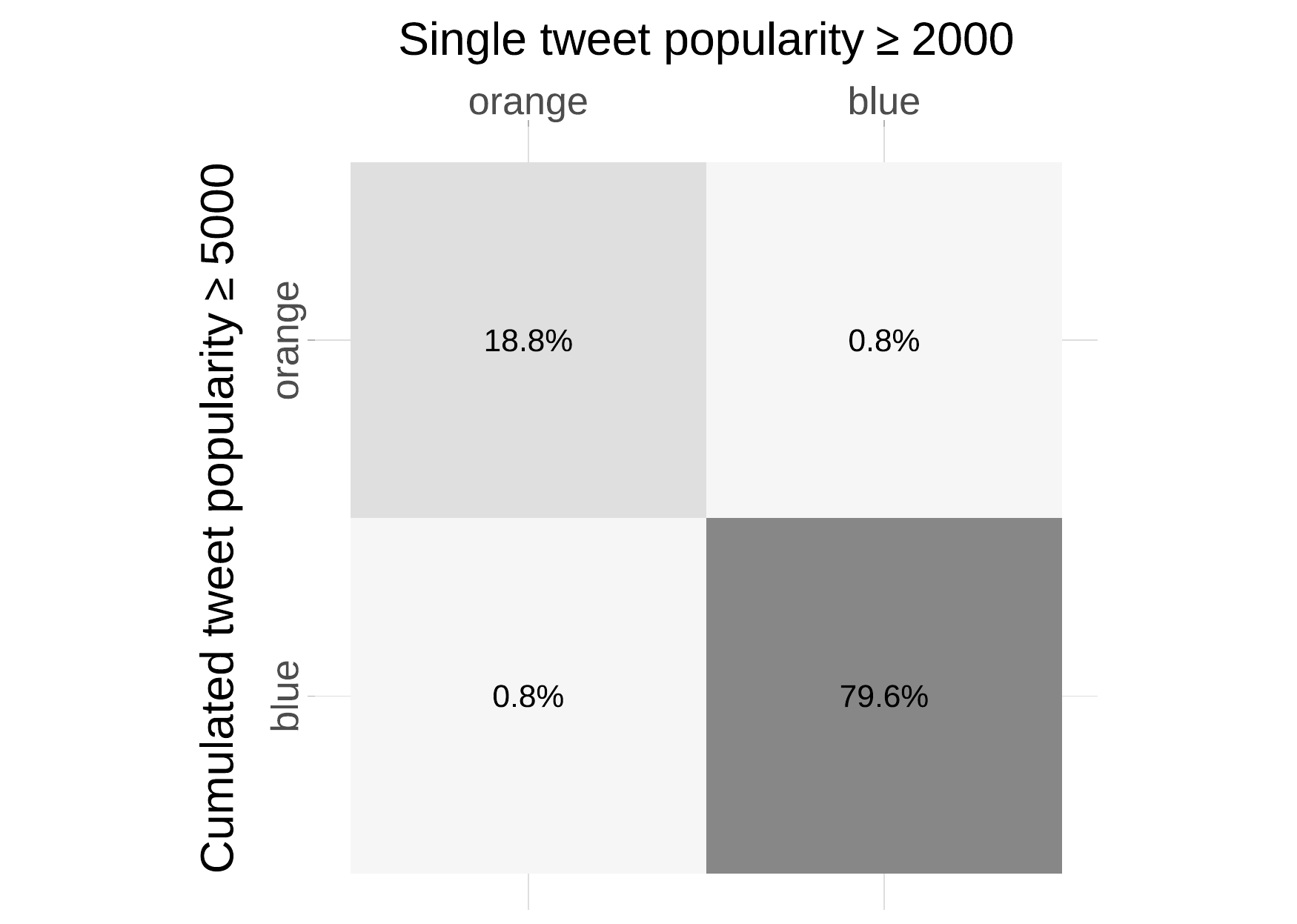}
         \caption{Confusion matrix for classifications in 
 Figure \ref{fig:total-pop} \\ versus main analysis (Figure \ref{fig:latentspace}).}
         \label{fig:confusion-total-pop}
     \end{subfigure}
\caption{
Comparison of user classifications obtained via the latent cluster random effects model using different popularity criteria. The confusion matrices included in the figure respectively compare classifications obtained using the networks in Figures \ref{fig:smaller}, \ref{fig:larger}, \ref{fig:atleast2}, \ref{fig:total-pop} against those obtained with the main network (Figure \ref{fig:latentspace}).} 
        \label{fig:confusion}
\end{figure}

To go beyond visual comparison, in addition to the plots we also constructed four confusion matrices. The matrices, shown in Figure \ref{fig:confusion}, compare the classification of users included in the main analysis with those obtained with each of the alternative elite networks. Note that not all users originally included in the elite network are present in the alternative networks due to the varying criterion and the resulting network size, but we can still compare the classification of users included in the main elite network which are still included in the alternative one. As an example, the first row of the matrix in Figure \ref{fig:confusion-smaller} can be interpreted as follows: 12.4\% of users were classified in the orange group in both the main analysis and the alternative one, while 0\% were classified as blue by the main analysis and as orange in the alternative one. Further, 1.5\% of users were classified as orange in the main and as blue in the alternative, while 86.1\% were classified as blue by both analyses. Looking at the four matrices, we can see that the vast majority of users that are present in both networks are classified in the same way, with only a few users switching communities with respect to the original labeling. Overall, these results thus demonstrate a good degree of robustness of our results to changes in the inclusion criterion, or, in other words, that the results are not very sensitive to the choice of elite definition being used. 

\section{Word frequency}
\begin{figure}
\centering \includegraphics[width=0.8\linewidth, trim=0cm 0cm 0cm 0cm, clip]{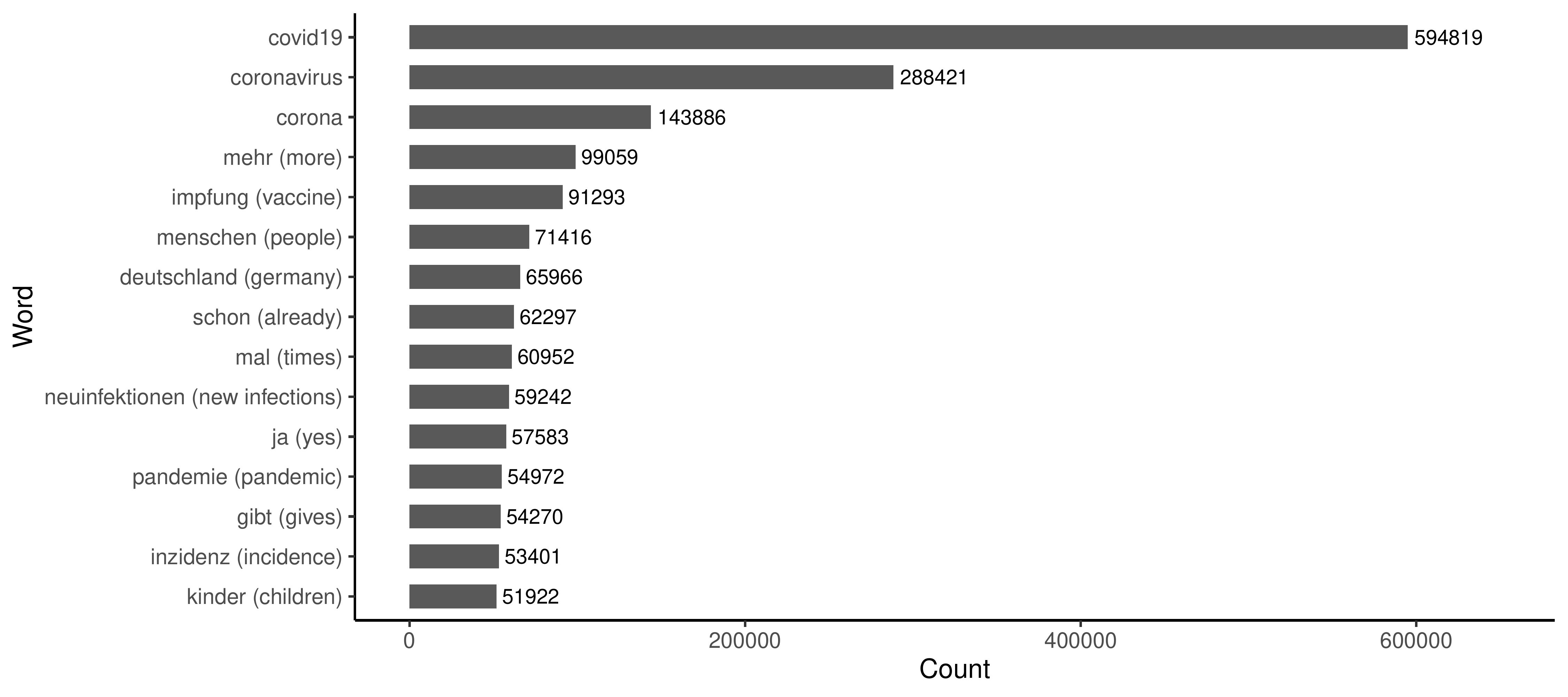} 

\caption{Frequency of the top 15 content words across the whole sample of tweets. The chart shows that vaccination was a central theme in the COVID-19-related Twitter discussion in 2021.}
\label{fig:wordcount}

\end{figure}

In the main body of the paper, we stated that vaccination was a prominent theme in the COVID-related discussion in 2021. To corroborate this, in Figure \ref{fig:wordcount} we provide a bar chart depicting the frequency of the top 15 content words across the complete sample. English translations of the words are included in parentheses where relevant. The table shows that, after ``covid19'', ``coronavirus'', ``corona'', and ``mehr'' (which means ``more''), the word ``impfung'' (meaning ``vaccine'') appears most often. Given that the word was not used to filter the dataset (see \cite{banda2021}), this gives an indication that vaccination was a central theme in the COVID-19-related Twitter conversation.

\newpage


\printbibliography

\end{refsection}

\end{document}